# Triple Halide Wide Bandgap Perovskites for Efficient Indoor Photovoltaics

Shaoyang Wang[1], Paul R. Edwards[2], Maged Abdelsamie[3], Peter Brown[4], David Webster[4], Arvydas Ruseckas[4], Gopika Rajan[5], Ana I. S. Neves[5], Robert W. Martin[2], Carolin M. Sutter-Fella[6], Graham A. Turnbull[4], Ifor D. W. Samuel[4] and Lethy Krishnan Jagadamma[1]*

[1] Energy Harvesting Research Group, School of Physics & Astronomy, SUPA, University of St Andrews, North Haugh, St Andrews, Fife KY16 9SS, United Kingdom

[2] Department of Physics, SUPA, University of Strathclyde, Glasgow G4 0NG, United Kingdom

[3] Materials Sciences Division, Lawrence Berkeley National Laboratory, California 94720, USA, United States

[4] Organic Semiconductor Centre, SUPA, School of Physics and Astronomy, University of St Andrews, North Haugh, St Andrews Fife, KY16 9SS, United Kingdom

[5] Department of Engineering, University of Exeter, North Park Road, EX4 4QF Exeter, United Kingdom

[6] Molecular Foundry, Lawrence Berkeley National Laboratory, California 94720, USA, United States

*Email: lkj2@st-andrews.ac.uk

**Abstract:** Indoor photovoltaics are receiving tremendous attention due to the continuous development of the Internet of Things (IoT). Here we report a triple anion (TA) perovskite $CH_3NH_3PbI_{2.6}(BrCl)_{0.2}$ with a tailored bandgap suitable for maximizing indoor light harvesting compared to methyl ammonium lead iodide $CH_3NH_3PbI_3$. The best-performing TA perovskite indoor-photovoltaic device achieved a steady-state power conversion efficiency (PCE) of 25.1% with an output power density of ~ 75 µW/cm$^2$ under 1000 lux indoor illumination (0.3 mW/cm$^2$ irradiance). This PCE is almost 40% higher than that of equivalent $CH_3NH_3PbI_3$-based devices (PCE of 17.9%). Longer carrier lifetime, reduced density of trap states and improved crystalline quality were achieved by the triple anion alloying method. The decisive role of chlorine (Cl) in the better performance of TA-based indoor photovoltaic devices was further investigated by successively reducing the Cl content and correlating it with the corresponding photovoltaic device performance. Replacing the commonly used hole transporting layer of Spiro-MeOTAD with undoped P3HT was found to significantly reduce the current-voltage hysteresis under indoor lighting conditions. A graphene-coated textile fiber-based temperature sensor was successfully powered by the triple anion perovskite indoor photovoltaic devices. The results from the present study demonstrate a novel route to maximize the PCE of halide perovskite indoor photovoltaic devices and their potential for application in the IoT industry.

## 1. Introduction

Indoor photovoltaic (indoor PV) technology is receiving rejuvenated research attention due to its potential for self-powering the innumerable wireless sensors in the huge technology field of the Internet of Things[1,2]. More than half of these wireless sensors are going to be inside buildings due to the anticipated radical changes in the built environment to realize smart and energy-secure buildings. The power requirements of the IoT components have been continuously decreasing in recent years: nowadays IoT wireless sensors need only a micro- to milliwatt range of power to operate, and efficient indoor photovoltaic cells are promising candidates to self-power them[1]. This autonomous powering of the IoT wireless sensors would reduce the dependence of this emerging technology on batteries and make it more environmentally sustainable and widely deployable. Among the various photovoltaic materials available today, hybrid halide perovskites are very promising for indoor light harvesting due to their various outstanding optoelectronic properties including tunable bandgap (≈1.2–3.1 eV)[3], high absorption coefficient (absorption length of 200–300 nm)[4], long carrier diffusion length (>1000 nm)[5] and high defect tolerance[6]. These promising properties have already triggered extensive research in perovskite solar cells and resulted in rapid development before their implementation in indoor PVs.

Developing efficient indoor PV starts with an understanding of the difference between indoor artificial light sources and outdoor sunlight. These light sources differ in spectrum and illumination intensity. As shown in **Figure 1(a)**, the illumination spectrum of modern indoor artificial light is much narrower than that of the sun. It emits only in the visible spectral range whereas the solar spectrum spans from the near- UV to mid IR. The visible emission of the indoor light source means that the optimal bandgap for a single junction solar cell is 1.9 eV (compared to 1.4 eV for 1 sun illumination) to maximize the PCE of indoor PV[7]. The wider bandgap of the photoactive layer can increase open circuit voltage ($V_{OC}$). In addition, the wider bandgap can also lead to the strong absorption of the narrow emission spectrum of the indoor light sources, increasing the short circuit current density ($J_{SC}$)[7,8]. A further important difference is that indoor light intensity is much lower than solar irradiance. The standard irradiance level for sunlight is defined as AM 1.5G which represents 100 mW/cm$^2$ (1 sun), while for indoor light, which is dominated by white light-emitting diodes (LEDs) and fluorescent lamps, the irradiance level is 0.05–0.5 mW/cm$^2$ and thus 100–1000 times lower than that of 1 sun illumination[9]. This dramatically lower light intensity makes defect control an important topic for perovskite-based indoor PVs since the beneficial effect of trap filling at the higher excitation density is no longer present. Therefore, there is greater possibility for trap-assisted recombination losses of the photogenerated charge carriers in the case of indoor PVs [10].

In ABX$_3$ perovskites, the valence band consists of a hybrid mixture of B site metal orbitals $n$s$^2$ and X site halide orbitals $n$p$^6$, with the major contribution from the latter. The conduction band is formed by a hybrid mixture of B site metal orbitals $n$p$^6$ and X site halide orbitals $n$p$^6$, with the major contribution from the former. The commonly used approach to widen the bandgap is to adjust the valence band-edge by compositional tuning of the halide ions. For the most widely investigated halide perovskite of methylammonium lead iodide (CH$_3$NH$_3$PbI$_3$), which has a bandgap of 1.56 eV, iodine ions locate at the halide (X-) site. When iodine ions are substituted by halides with lower energy $p$ orbitals, such as bromine ion and chlorine ion, the valence band is lowered significantly by 0.60 eV[11]. This makes mixed halide perovskite composition tuning a viable route to widen the bandgap in order to maximize the indoor light harvesting properties[8,10,12,13]. Following this approach, iodide-bromide alloying has successfully shifted the bandgap from 1.6 to 1.75 eV with a 40% Br ratio[14]. However, it is reported that a larger amount of Br incorporation (greater than 20% halide mole fraction) into CH$_3$NH$_3$PbI$_3$ results in phase segregation under illumination or during aging[15]. Upon phase segregation, the perovskite phase is transformed into I-rich and Br-rich domains; the excited electrons will relax down to the lower bandgap of the I-rich domains, making the bandgap shift ineffective and reducing the $V_{OC}$. Hence, the addition of Cl to give the triple anion (TA) system is presented as an effective method to suppress phase segregation by modifying the morphology and surface passivation and to realize efficient bandgap tuning[16].

The wide bandgap TA perovskites have been successfully used in perovskite-silicon tandem solar cells[16]. However, little is known yet regarding their suitability for indoor photovoltaic applications. Our recent study on halide perovskite indoor PV has demonstrated the necessity of optimizing the device architecture and photoactive layers separately for indoor and 1 sun illumination conditions[17]. Kim *et al.* found that mixed cation-triple anion perovskite [$FA_{0.963}MA_{0.037}PbI_{2.813}Br_{0.037}Cl_{0.15}$] is beneficial for suppressing ion migration and non-radiative recombination and reported a PCE of ~20% and output power density of 35.25 µW/cm$^2$ (based on transient *J-V* measurements) under 800 lux LED illumination[10]. Cheng *et al.* reported that triple-anion perovskite ($CH_3NH_3PbI_{2-x}BrCl_x$) can have enhanced charge carrier lifetime and suppressed light-induced phase segregation up to 100 suns compared to $CH_3NH_3PbI_3$ and $CH_3NH_3PbI_2Br$, reporting a higher PCE of 36.2% under 1000 lux fluorescent light (275.4 µW/cm$^2$)[8]. These results indicate that triple anion alloying could be a promising method to develop efficient indoor PVs[8,10,18]. However, the above reports on triple anion indoor PVs employ a two-step fabrication process for the incorporation of chloride ions ($CH_3NH_3Cl$, $CH_3NH_3Br$ or $HC(NH_2)_2I$ deposition after the $CH_3NH_3PbI_3$ film formation) which increases the complexity of device fabrication, and they lack discussion of *J-V* hysteresis and the steady-state power output from these devices. Our recent study has shown that, under indoor lighting conditions, the *J-V* hysteresis effects can become more significant than for 1 sun illumination and hence steady-state power output measurements should be prioritized over the conventional transient *J-V* scan[17].

In the present work, we report a facile single-step fabrication of TA halide perovskite composition of $CH_3NH_3PbI_{2.6}(BrCl)_{0.2}$ with a tailored bandgap of 1.69 eV and compare its indoor light harvesting and *J-V* hysteresis properties with that of $CH_3NH_3PbI_3$. To gain more insight into the benefits of the triple anion alloying method in indoor photovoltaics, we systematically investigated the microstructural, photophysical and optoelectronic properties of $CH_3NH_3PbI_{2.6}(BrCl)_{0.2}$ films, partial heterostructures and completed devices and compared them to those of $CH_3NH_3PbI_3$. Our study revealed that the triple anion composition of $CH_3NH_3PbI_{2.6}(BrCl)_{0.2}$ suffers significantly fewer defect-related recombination losses, exhibits enhanced charge carrier lifetime, and possesses better crystalline properties in comparison to CH$_3$NH$_3$PbI$_3$ resulting in improved power conversion efficiency and suppressed hysteresis effects under indoor lighting conditions. The decisive role of Cl in the better performance of CH$_3$NH$_3$PbI$_{2.6}$(BrCl)$_{0.2}$ based indoor photovoltaic devices is further verified by successively reducing the Cl content and correlating it with the corresponding photovoltaic device performance.

## 2. Results and Discussion

### 2.1. Microstructural characterization of triple anion perovskite films

To obtain triple anion CH$_3$NH$_3$PbI$_{2.6}$(BrCl)$_{0.2}$, equimolar amounts of PbBr$_2$ and PbCl$_2$ were added into the PbI$_2$ stoichiometry-adjusted CH$_3$NH$_3$PbI$_x$ precursor solution (with the nominal perovskite composition then assumed to follow that of the precursor stoichiometry). The standard CH$_3$NH$_3$PbI$_3$ perovskite is used as the control sample in the study. During spin coating, both films were washed by anti-solvent diethyl ether followed by a thermal annealing treatment for 2 minutes on a hotplate at 100 °C. A detailed description of the preparation method is given in the experimental section in the supporting information. The UV-vis spectroscopy measurement was performed initially on TA films and CH$_3$NH$_3$PbI$_3$ control samples to characterize the bandgap properties. As shown in **Figure 1 (b)** the absorption edge for the TA perovskite composition is blue-shifted from ~775 nm (CH$_3$NH$_3$PbI$_3$) to 750 nm (TA). This corresponds to an increase of the bandgap from 1.61 eV for CH$_3$NH$_3$PbI$_3$ to 1.69 eV for the TA films (Figure S1). The valence band (VB) position was characterized by ambient photoemission spectroscopy (APS) as shown in **Figure 1 (c).** The TA films showed a VB edge of 5.46 eV, deeper than that of the CH$_3$NH$_3$PbI$_3$ VB at 5.39 eV as expected due to the incorporation of Br$^-$ and Cl$^-$ anions.

To understand the crystalline properties, X-ray diffraction (XRD) characterization was performed (Figure S2). Peaks at 14.18° and 28.6° can be indexed to the (110) and (220) diffraction peaks of the tetragonal perovskite phase. For both diffraction peaks, the TA sample shows significantly higher peak intensity indicating its enhanced crystallinity compared to CH$_3$NH$_3$PbI$_3$. Also, the TA sample revealed the existence of PbI$_2$ with the appearance of a small peak at 12.65°. This observation of PbI$_2$ agrees with the previous research on perovskite compositions containing bromide and chloride anions[16,19]. **Figure 1 (d)**, shows a typical peak shifting from 14.11° for CH$_3$NH$_3$PbI$_3$ to 14.20° for the TA sample as expected due to shrinkage of the perovskite lattice due to incorporation of $Cl^-$ (1.81 Å) and $Br^-$ (1.96 Å) with smaller ionic radii compared to $I^-$ (2.20 Å).[8] In addition, a new peak is observed at 15.49°, which is considered to be a characteristic peak of $CH_3NH_3PbCl_3$ as per previous reports[16,19,20]. The main challenge with the triple anion alloying compared to the iodide-bromide double halide system is to confirm the presence of Cl in the perovskite lattice, because it can volatilize during the thermal annealing process as $CH_3NH_3Cl$[18]. The existence of the $CH_3NH_3PbCl_3$ peak provides primary proof that Cl is incorporated and remains within the perovskite active layer instead of volatilizing. Previous research has pointed out the difficulty in detecting chlorine within the perovskite layer after the thermal annealing process although it was present in the precursor solution[21]. In the present study, the existence of chlorine in the $CH_3NH_3PbI_{2.6}(BrCl)_{0.2}$ films were further confirmed using wavelength dispersive X-ray (WDX) spectroscopy. By contrasting the WDX results from the $CH_3NH_3PbI_3$ and TA samples respectively, it is evident that Cl is incorporated within the TA samples. **Figure 1 (e)** compares the WDX counts from the Cl content of $CH_3NH_3PbI_3$ and TA samples; the $CH_3NH_3PbI_3$ sample has a bremsstrahlung (continuum) background of ~300 counts, while the TA sample shows a clear characteristic X-ray peak (~1600 counts) at the Cl $K_\alpha$ energy, confirming the presence of Cl within the perovskite active layer. Monitoring the WDX peak count rates over 10 minutes verified that there was minimal dissociation/volatilisation caused by the incident 8 keV electron beam [Figure S3 (a–c)]. This precaution was taken based on the previous studies, where it has been shown that under the high dose electron beam, methylamine and hydrogen iodide (hydrogen halide) can escape from the halide perovskite samples due to electron beam induced damage[22–24]. **Figure 1 (f)** shows the SEM images of $CH_3NH_3PbI_3$ and TA thin film samples. For both films, the surface morphology was compact and dense whereas, in the case of the TA sample, dark and bright contrast domains were observed. According to previous research, the brighter domains may be $PbI_2$[25]. Although the XRD spectra of the TA samples indicated the presence of $PbI_2$, its presence has been further verified using low electron beam voltage (and hence surface-sensitive) cathodoluminescence (CL) spectroscopy. As shown in **Figure 1(g)**, the CL emission spectrum of the TA sample shows two emission peaks; one corresponding to its near band-edge emission and the other at ~2.5 eV arising from the $PbI_2$[25]. Also, the near band-edge CL emission of the TA sample is blue-shifted compared to the control $CH_3NH_3PbI_3$ sample as expected. The peak of the near band-edge CL emission energy from both the $CH_3NH_3PbI_3$ (1.63 eV) and TA (1.73 eV) sample is found to be slightly higher than the band gap energy estimated using the UV-vis absorption spectra. This could be due to the bandgap energy estimated from the absorption edge being lower than the actual bandgap due to the existence of tail states in the CH$_3$NH$_3$PbI$_3$ and TA films[26]. Also, the UV-vis absorption measurements probe the full thickness (350 nm) of the sample, whereas the CL analysis depth is limited by the beam energy; with the 5 keV electron beam used, the penetration depth and hence the signal generation depth is estimated as ≈ 240 nm using a Monte-Carlo simulation method[27]. This would mean that any vertical compositional heterogeneity, as the authors have previously noted in all-inorganic mixed halide perovskites[28], can also contribute to this slightly higher CL emission energy compared to the band gap energy estimated from the UV-vis measurements.

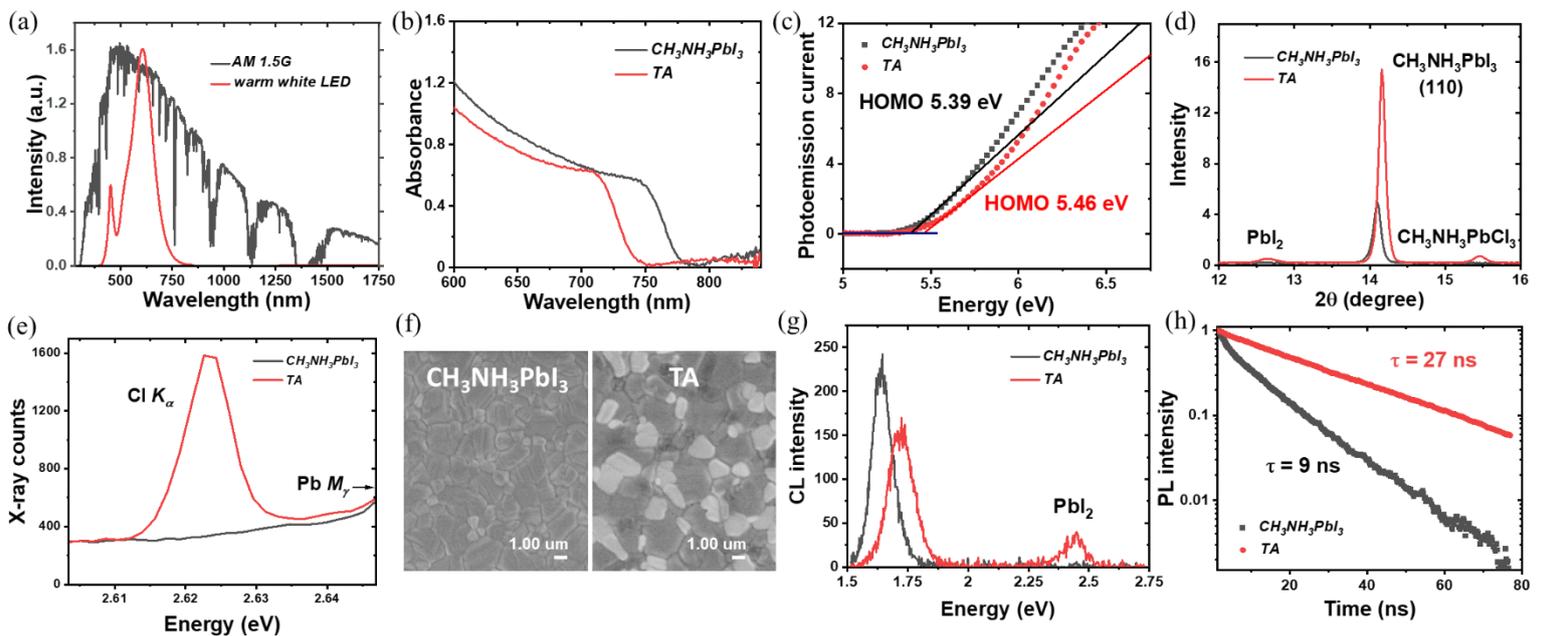

*Figure 1. (a) Comparison of the 1 sun spectrum with the warm white LED indoor light source used in this study (b) Absorbance spectra of triple anion perovskite film and CH$_3$NH$_3$PbI$_3$ control film from UV-Vis spectroscopy. (c) Ambient photoemission spectra of TA and CH$_3$NH$_3$PbI$_3$ films. (d) X-ray diffraction pattern of TA and CH$_3$NH$_3$PbI$_3$ films (e) Wavelength-dispersive X-ray spectra of TA and CH$_3$NH$_3$PbI$_3$ films in the region of the Cl K$_α$ X-ray line showing incorporation of Cl in the TA film. (f) Scanning electron microscopy images of TA and CH$_3$NH$_3$PbI$_3$ films. (g) Room temperature CL spectra of TA and CH$_3$NH$_3$PbI$_3$ films showing the blue-shifted near bandage emission from the TA films and the appearance of PbI$_2$ emission. (h) Time-resolved photoluminescence spectra of TA and CH$_3$NH$_3$PbI$_3$ films. Excitation was at 515 nm*

The PbI$_2$ induced by the triple anion alloying method and the incorporated Cl has been reported to effectively passivate the defects in the perovskite layer and reduce non-radiative recombination[25,29–31]. We further tested this in our samples by making measurements of time-resolved photoluminescence (TRPL) measurements shown in **Figure 1(h)**. For the TRPL measurements, the perovskite layers were deposited onto bare ITO substrates. The measured TRPL data clearly show a much slower decay for TA films than for CH$_3$NH$_3$PbI$_3$ films. The fitted decay time $\tau$ is 27 ns for TA films and 9 ns for CH$_3$NH$_3$PbI$_3$ control films, indicating fewer trap states and non-radiative recombination losses for triple anion films. After confirming the wider bandgap, better crystalline quality, compact surface morphology, enhanced PL lifetime and the presence of Cl in the CH$_3$NH$_3$PbI$_{2.6}$(BrCl)$_{0.2}$ films, the photovoltaic properties of the TA films were characterized and compared to those of CH$_3$NH$_3$PbI$_3$.

## 2.2. Photovoltaic properties

To investigate the photovoltaic device performance of the TA perovskites, devices were fabricated in typical n-i-p planar architecture with a layer structure of glass/ITO/SnO$_2$/perovskite/Spiro-OMeTAD/Au. The *J-V* scan measurements were performed for the CH$_3$NH$_3$PbI$_3$ and TA devices under indoor warm white LED illumination of 1000 lux (0.3 mW/cm$^2$) and 1 sun. The *J-V* characteristics and the PCE distribution from these measurements are shown in **Figure 2**. The corresponding photovoltaic performance parameters are shown in Figure S4, **Table 1** and Table S1. The box plots in Figures 2 and S4 present the distribution and average PCE of more than 20 photovoltaic devices and performance parameters of $V_{OC}$, FF and $J_{SC}$. The *J-V* characteristics of the champion devices under indoor illumination are given in **Figure 2(a)**. The TA device shows a maximum PCE of 26.1% for the forward scan and 33.6% for the reverse scan. The CH$_3$NH$_3$PbI$_3$ control devices present a forward scan PCE of 21.0% and reverse scan PCE of 30.1% for their best performance. As shown in the box plots in **Figure 2(b)** and Table 1, the average PCE of TA devices reaches 22.6% for the forward scan and 30.1% for the reverse scan while for CH$_3$NH$_3$PbI$_3$ devices it is only 17.8% and 27.8% respectively, which shows a substantial enhancement of indoor light harvesting by the TA films. From **Figure 2 (a) & (c)**, and Table 1 it can be seen that a significant part of the enhancement in PCE of the TA device under indoor lighting is due to the higher $V_{OC}$ which is 0.86 V and 0.89 V for different scan directions. The corresponding $V_{OC}$ for the CH$_3$NH$_3$PbI$_3$ control sample is only 0.78 V and 0.84 V, respectively. The fill factor of TA devices is also consistently improved particularly for the forward scan, the average FF of TA devices is 57.2% while that of $CH_3NH_3PbI_3$ devices is only 46.5%. The significant improvement of FF indicates better charge extraction and collection in the TA films and can attribute to the better crystalline quality of TA films compared to $CH_3NH_3PbI_3$. On the other hand, the $J_{SC}$ of both types of devices are comparable under indoor illumination as shown in Figure S4 (c) & Table 1 (forward scan: 0.15 mA/cm$^2$ for $CH_3NH_3PbI_3$ vs 0.14 mA/cm$^2$ for TA; reverse scan: 0.14 mA/cm$^2$ for $CH_3NH_3PbI_3$ vs 0.13 mA/cm$^2$ for TA). **Figure 2 (f)** shows the external quantum efficiency (EQE) spectra. The blue shift of the TA absorption edge, compared to $CH_3NH_3PbI_3$ is evident and supports the UV-vis absorption measurements. **Figure 2(f)** also shows the emission spectrum of the indoor warm white LED source used in the present study to explore the indoor photovoltaic properties. The larger bandgap of TA composition reduces the thermalization losses under the indoor light illumination, resulting in higher $V_{OC}$ without sacrificing $J_{SC}$.

As for the device performance under 1 sun, from the *J-V* curves in **Figure 2 (d)**, TA devices reach a maximum PCE of 14.6% for forward scan, and 16.5% for reverse scan, while the PCE of the $CH_3NH_3PbI_3$ control device is only 12.8% for the forward scan and 13.4% for the reverse scan for its champion device. From **Figure 2 (e),** it is noticed that the overall PCE of TA devices is slightly improved compared to $CH_3NH_3PbI_3$ under 1 sun. A comparison of the photovoltaic performance parameters in Figure S4 (d) shows that the average $V_{OC}$ of TA devices is improved to 1.06 V compared to that of 0.9 V for $CH_3NH_3PbI_3$. This observation of higher $V_{OC}$ for TA devices is consistent with their larger bandgap compared to $CH_3NH_3PbI_3$, as revealed by UV-vis spectroscopy and APS measurements [Figure 1 (b) & (c)].

The $J_{SC}$ from both types of devices are broadly comparable (forward scan: 19.0 mA/cm$^2$ for TA vs 19.2 mA/cm$^2$ for $CH_3NH_3PbI_3$; reverse scan: 17.9 mA/cm$^2$ for TA vs 18.0 mA/cm$^2$ for $CH_3NH_3PbI_3$ from Figure S4 (f) and Table S1). The fill factors of these devices are also within a similar range, being 61.3% and 56.9% for forward, and 70.1% and 71.6% for reverse scans for $CH_3NH_3PbI_3$ and TA devices, respectively. These results show that the triple anion alloying method is effective in boosting the $V_{OC}$ without compromising $J_{SC}$ and FF thus yielding higher PCE.

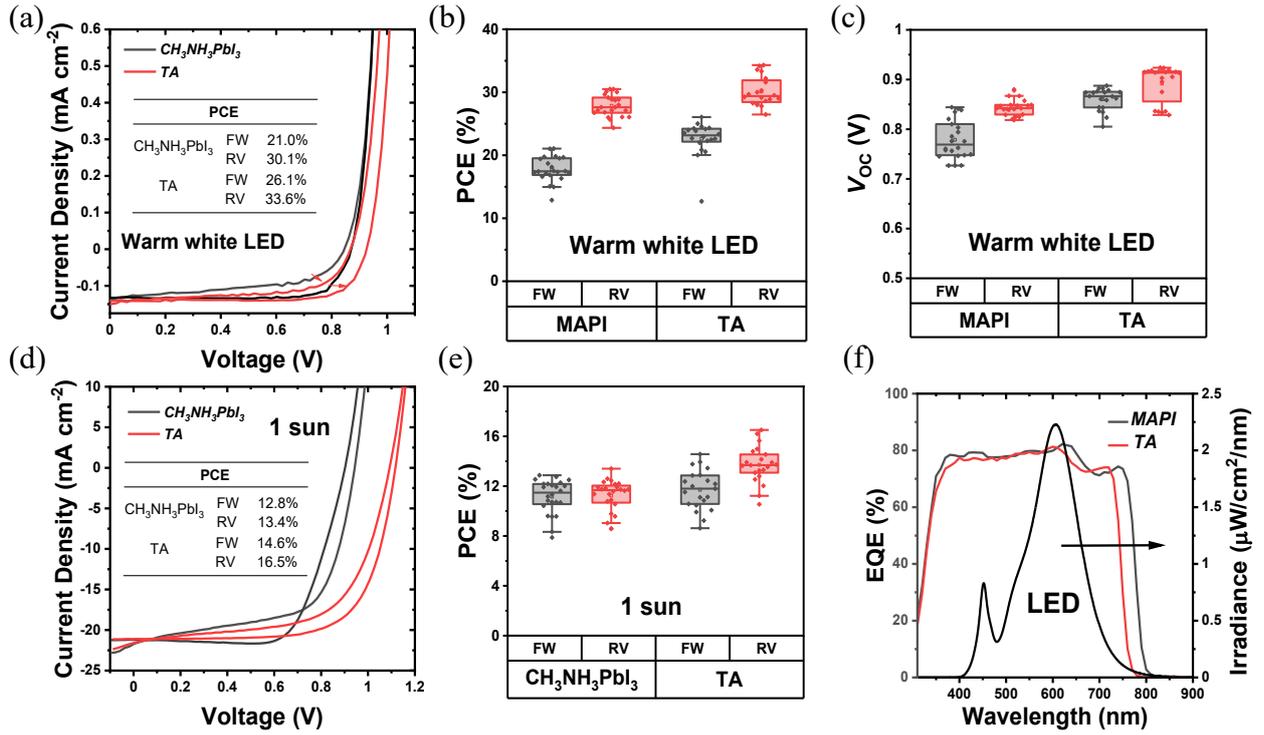

*Figure 2. (a) J-V curves of TA and CH$_3$NH$_3$PbI$_3$ based devices under warm white LED illumination. (b) The statistical distribution of PCE values of TA and CH$_3$NH$_3$PbI$_3$ devices under warm white LED illumination. (c) The distribution of $V_{OC}$ of two types of photovoltaic devices under warm white LED illumination. (d) J-V curves of TA and CH$_3$NH$_3$PbI$_3$ under 1 sun illumination. (e) The statistical distribution of PCE values of TA and CH$_3$NH$_3$PbI$_3$ devices under 1 sun illumination. (f) EQE spectra of the TA and CH$_3$NH$_3$PbI$_3$ devices. The irradiance spectrum of the warm white LED illumination used in the present study is also shown.*

**Table 1.** *Photovoltaic performance parameters of CH$_3$NH$_3$PbI$_3$ and TA device under 1000 lux warm white LED illumination*

| Device type | | Average PCE (%) | Average FF (%) | Average Jsc (mA/cm$^2$) | Average $V_{oc}$ (V) |
|---|---|---|---|---|---|
| CH$_3$NH$_3$PbI$_3$ | FW | 17.8 ± 2.0 | 46.5 ± 4.1 | 0.147 ± 0.004 | 0.779 ± 0.039 |
| | RV | 27.8 ± 1.7 | 75.4 ± 3.7 | 0.131 ± 0.004 | 0.844 ± 0.017 |
| TA | FW | 22.6 ± 2.8 | 57.2 ± 5.6 | 0.138 ± 0.009 | 0.860 ± 0.022 |
| | RV | 30.1 ± 2.3 | 78.4 ± 5.0 | 0.129 ± 0.007 | 0.892 ± 0.036 |

Since *J-V* hysteresis exists for the photovoltaic devices under both 1 sun and indoor illuminations, the PCE of the devices was also measured using the steady-state method of maximum power point tracking (MPPT). Steady-state MPPT measurement holds the device at the highest power output point for a period, which simulates the scenario where the photovoltaic device used in the real industrial or domestic application of powering the load. The PCE obtained from the MPPT measurements is shown in **Figure 3 (a)**. Under 1000 lux warm white LED illumination, the steady-state PCE of TA device reaches as high as 25.1%, which is significantly higher than 17.9% for $CH_3NH_3PbI_3$ devices. Thus under 1000 lux illuminance conditions, the TA devices can deliver a steady-state power output of 75.4 $\mu W/cm^2$ compared to the 53.7 $\mu W/cm^2$ from the $CH_3NH_3PbI_3$ devices. These results are summarized in **Table 2**. Regarding the current IoT wireless protocols industry, the 75.4 $\mu W/cm^2$ power density would allow a 20 $cm^2$ TA perovskite indoor photovoltaic device to power most of the RFID, LoTA Backscatter, Passive Wi-Fi, BLE, ANT and ZigBee nodes[2]. The excellent steady-state PCE of the TA device further proves the superiority of the triple anion alloying method. The steady-state PCE of the devices under 1 sun illumination condition is shown in **Figure 3 (b)**; the steady-state PCE of TA devices is 13.5%, slightly higher (13.1%) than that of $CH_3NH_3PbI_3$, which still emphasizes the better quality of TA devices.

**Table 2.** *Summary of steady state PCE and output power density for $CH_3NH_3PbI_3$ and TA devices*

| Light source | Device type | Steady state PCE (%) | Output Power Density ($\mu W/cm^2$) |
|---|---|---|---|
| Warm white LED (1000 lux) | $CH_3NH_3PbI_3$ | 17.9 | 53.7 |
| | TA | 25.1 | 75.4 |

To gain more insight into the enhanced photovoltaic properties of TA devices, light intensity-dependent *J-V* characterization, transient photocurrent (TPC) and photovoltage (TPV) measurements were carried out. Under the low-intensity indoor lighting conditions suppression of trap states is very crucial to maximize the power output and therefore non-radiative recombination losses were studied by plotting $V_{OC}$ vs light intensity $(L)$[32]. The relationship between light intensity $(L)$ and $V_{OC}$ is given as[33]:

$$V_{OC} = \frac{n_{id}k_BT}{q} \ln\left(\frac{I_L}{I_0} + 1\right) = \frac{n_{id}k_BT}{q} \ln L, \quad (1)$$

where $k_B$ is the Boltzmann factor, $n_{id}$ is an ideality factor, $I_L$ is the total solar cell current under illumination (photocurrent), and $I_0$ is the dark saturation current. Since the current under illumination is much higher than the dark current and photocurrent is linearly related to light intensity, $L$ can be approximated as the ratio of $I_L$ and $I_0$.

In general, an ideality factor of 1 (or close to 1) indicates the dominant recombination mechanism is bimolecular (radiative) recombination, whereas values closer to 2 indicate pronounced trap-assisted Shockley-Read-Hall (SRH) recombination[34,35]. From **Figure 3(c)**, the ideality factor calculated for $CH_3NH_3PbI_3$ is 2.03. while for TA, it is only 1.41, indicating that the non-radiative recombination is significantly reduced for the TA devices, which is consistent with the TRPL measurement shown in Figure 1(h). In addition to TRPL characterization of TA and $CH_3NH_3PbI_3$ films on bare ITO substrates, TRPL was also carried out for perovskite films deposited on $SnO_2$/ITO films ($SnO_2$ is the ETL in the device architecture considered here). By taking the TRPL ratio of perovskite films on electron transport layer $SnO_2$ to the perovskite films on bare ITO substrate, we account for the natural PL decay in perovskite films and get the decay which is caused solely by electron extraction. The PL ratio shows very similar decay time of ~40 ns for both perovskites which indicates that the electron extraction rate is the same for $CH_3NH_3PbI_3$ and TA (Figure S5)[36]. Hence, the longer lifetime of free carriers from TA perovskite film provides more efficient charge extraction which significantly enhances FF. TPV and TPC measurements were used to further investigate the improvements of carrier lifetime. The advantage of these measurements is that they can be made on completed devices under the conditions of practical photovoltaic operation. **Figure 3(d)** shows carrier lifetime variation as a function of $V_{OC}$ (light intensity). The steep decrease of carrier lifetime at higher $V_{OC}$ is attributed to faster carrier recombination owing to higher light intensity. The results in Figure 3(d) reveal that the carrier lifetime of TA devices is consistently higher than that of $CH_3NH_3PbI_3$ devices, supporting their improved photovoltaic performance as shown in Figure 2, and implying their reduced trap states and better interface conditions. The equation of carrier lifetime versus $V_{OC}$ used to extract $\beta$ is given as below (extracted from TPV measurements) [37]:

$$\tau = \tau_0 e^{-\beta V_{OC}}, \quad (2)$$

where $\tau$ is the carrier lifetime and $\beta$ is the decay constant obtained by fitting the TPV data. From **Figure 3(d)**, $\beta$ is 17.7 μs V$^{-1}$ and 12.2 μs V$^{-1}$ for TA and CH$_3$NH$_3$PbI$_3$ devices, respectively. The enhanced voltage decay constant of TA devices also suggests a longer carrier lifetime and better device quality. **Figure 3(e)** shows the TPC charge extraction results for the devices. The equation used for the fitting is:

$$n = n_0 e^{\gamma V_{OC}}, \quad (3)$$

TA devices exhibit a higher $\gamma$ parameter (10.5 cm$^{-3}$V$^{-1}$) for the rate of increase of charge density $n$ compared to that of CH$_3$NH$_3$PbI$_3$ (8.4 cm$^{-3}$V$^{-1}$). The $\gamma$ parameter is the rate of increase of $n$ with respect to bias which has a value of 19 V$^{-1}$ for an ideal semiconductor ($\gamma \approx e/2k_BT$)[37]. The larger deviation from ideality of the $\gamma$ parameter of the CH$_3$NH$_3$PbI$_3$ device indicates more non-radiative recombination, suggesting a higher density trap states for CH$_3$NH$_3$PbI$_3$ devices compared to TA devices, and agrees with the findings of $V_{OC}$ vs light intensity measurements. The higher $\gamma$ parameter of TA devices implies better interfacial conditions which also agrees with the improved decay lifetime $\tau$ from the TRPL measurements. The perturbed lifetime $\tau_{\Delta n}$ can be related to the total charge density with a power dependence in η through the following equation[37]:

$$\tau_{\Delta n} = \tau_{\Delta n_0} \left(\frac{n_0}{n}\right)^{\eta}, \quad (4)$$

The exponent $\eta$ can be obtained by fitting $\tau_{\Delta n}$ vs the charge density $n$; it can also be obtained by rearranging equation (2) and (3) ($\eta = \beta/\gamma$). $\eta$ for TA and CH$_3$NH$_3$PbI$_3$ devices is 1.68 and 1.44, respectively. The $\eta$ exponent can be used to calculate the total lifetime for devices from the perturbed lifetime using equation (5). With the higher $\eta$ value, the calculated total lifetime will be higher for TA devices with higher intrinsic perturbed lifetime, which further confirms reduced trap states from triple anion perovskites.

$$\tau_n = (\eta + 1)\tau_{\Delta n}. \quad (5)$$

The reduced non-radiative recombination, prolonged lifetime and better interfacial properties can be linked to the improved device performance and emphasize the importance of controlling trap for maximizing the efficiency of indoor photovoltaic devices.

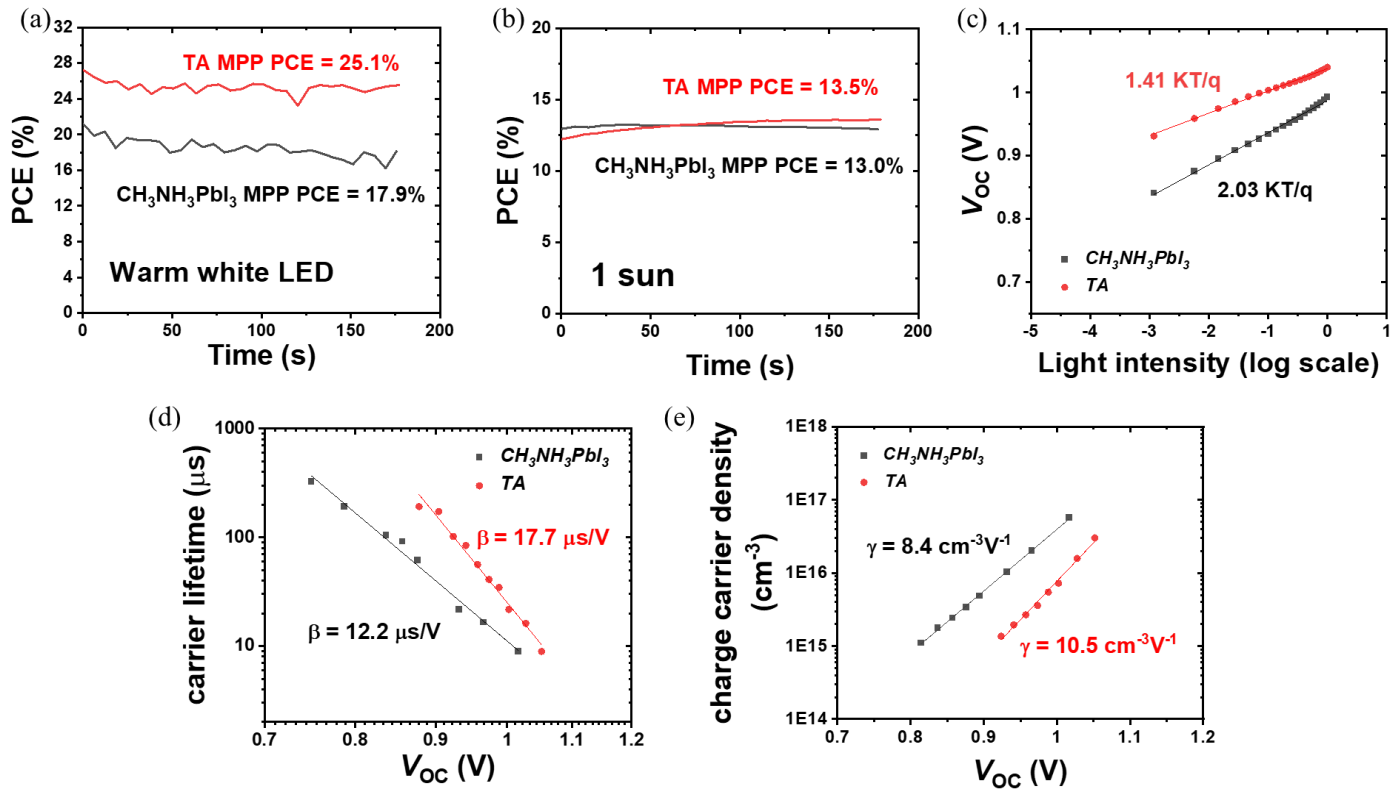

*Figure 3. (a) MPPT PCE comparison of TA and $CH_3NH_3PbI_3$ devices under warm white LED illumination. (b) MPPT PCE comparison of TA and $CH_3NH_3PbI_3$ devices 1 sun illumination. (c) $V_{OC}$ variation versus light intensity. (d) Transient photovoltage characterization of TA and $CH_3NH_3PbI_3$ devices. (e) Transient photocurrent characterization of TA and $CH_3NH_3PbI_3$ devices.*

## 2.3. Role of Cl in the enhanced photovoltaic properties of TA devices

To obtain deeper insight into the better indoor photovoltaic properties of the TA composition, the influence of the halide content needs to be investigated. Since bromine is relatively stable, it is particularly important to investigate how the photovoltaic device properties are influenced by chlorine content. Previously, thermal annealing process has been successfully used to vary the Cl content in a double halide $CH_3NH_3PbI_{3-x}Cl_x$ perovskite layer[19]. This was made possible because of the volatile nature of Cl, allowing the incorporated chlorine to be released in the form of $CH_3NH_3Cl$ during thermal annealing. In the present investigation, we used a similar thermal annealing method to control the content of chlorine in the TA perovskite layer. We selected the thermal annealing steps of 2 min, 10 min, 30 min, 45 min and 60 min with the same temperature 100 °C to investigate the effect of chlorine content. We used WDX spectroscopy to estimate the Cl content in the resultant TA perovskite layers. The iodine and bromine contents are relatively stable during the different duration of the thermal annealing process as shown in Figure S6. **Figure 4 (a)** shows WDX spectra in the region of the chlorine $K_\alpha$ characteristic X-ray line, after background correcting by subtracting the corresponding spectrum for Cl-free $CH_3NH_3PbI_3$. This background correction accurately accounts for both the bremsstrahlung continuum radiation and the tail of the nearby Pb $M_\gamma$ line at 2.653 keV (see **Fig. 2e**). The resultant net X-ray peak intensity shows the trend of chlorine content decline on increasing the annealing duration. The 2 min-annealed sample has the highest chlorine content with a peak of over 1200 X-ray counts, reducing to a near-negligible signal after 60 min annealing. In the absence of readily available WDX standards of a suitably close composition, it is adequate for the purposes of this work to estimate relative compositions based on the assumptions that (i) the 2-minute annealed TA sample is close to its nominal $CH_3NH_3PbI_{2.6}Br_{0.2}Cl_{0.2}$ stoichiometry, and (ii) any atomic number, X-ray absorption and secondary fluorescence ("$ZAF$") effects are minimal. The estimated Cl wt % is thus 1.2%, 0.74 %, 0.13 %, 0.03 % and 0.003% respectively for the 2-, 10-, 30-, 45- and 60-minute annealed TA samples. This WDX result matches the results from the Sun *et al.* study, which constructs a Cl content gradient to investigate the effect of chlorine[19]. It is noteworthy that, except for the 2 minute thermally annealed sample, the Cl content in different TA samples is relatively stable with respect to the WDX electron beam irradiation duration of 10 minutes, as shown in Figure S6(f). While some slight reduction in counts is seen in the Cl peak intensity (strongest, as expected, in the 2 minutes thermally annealed TA sample which retains the most Cl), the magnitude is not sufficient to have a significant effect on the measured WDX compositions.

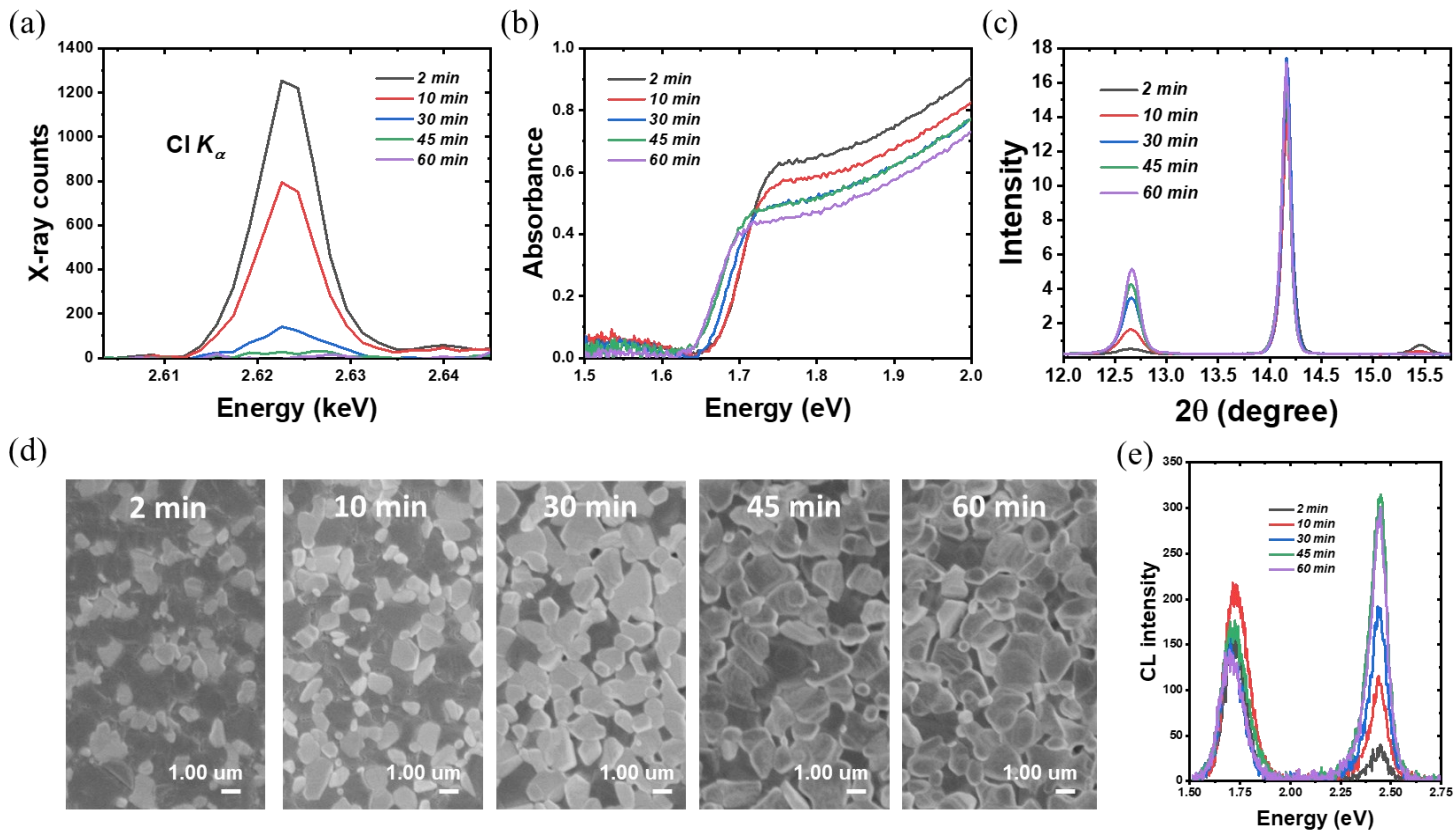

*Figure 4. (a) Background-corrected wavelength dispersive X-ray spectra of the Cl $K_\alpha$ line as a function of thermal annealing time. (b) Absorbance spectra of triple anion incorporated perovskite film based on different thermal annealing times. (c) X-ray diffraction pattern showing the effect of thermal annealing of the TA films. (d) secondary electron (SE) images of annealed films as a function of different thermal annealing duration. The contrast domains represent $PbI_2$. (e) Cathodoluminescence spectra of TA films showing the enhanced CL emission with the increase in thermal annealing duration.*

**Figure 4 (b)** shows the UV-vis absorption spectra of the TA films with different Cl content, revealing a redshift with the decreasing chlorine content. The absorption edge of the 2 min annealed sample is at ~1.69 eV and is gradually shifted to ~1.64 eV for the 60 min annealed sample, in which the chlorine content is negligibly small. The redshift in the absorption edge with a longer annealing time is consistent with the WDX characterization that chlorine is constantly reduced in the thermal annealing process. The crystalline properties of the TA perovskite layer as a function of thermal annealing (and hence as a function of chlorine content) were investigated by XRD. The peaks at 14.18° and 28.6° which index to (110) and (220) planes of the tetragonal perovskite phase remain. Another significant change can be noticed from the $CH_3NH_3PbCl_3$ characteristic peak at 15.49°. Though the $CH_3NH_3PbCl_3$ peak can be identified from the 2 min-annealed samples (with the highest chlorine content as per the WDX data), it is then reduced dramatically in the 10 min-annealed sample and completely disappeared in other samples, which further implies the escape of chlorine during the thermal annealing process. In addition, along with decreasing chlorine, the intensity of the $PbI_2$ characteristic peak at 12.65° is enhanced as the thermal annealing time increases.

**Figure 4(d)** shows the secondary electron SEM images of the TA films as a function of different thermal annealing duration. From **Figure 4 (d)**, with longer thermal annealing time, the density of white domain-like features most likely related to ($PbI_2$ phase) is increased, which is consistent with the XRD results. For the 45 min and 60 min annealed samples, which only have a trace amount of chlorine, the $PbI_2$ features dominate the film surface. The increased $PbI_2$ further matches the CL emission results shown in **Figure 4 (e)**. Besides the TA perovskite near band-edge emission peak, the $PbI_2$ peak at 2.4 eV is significantly enhanced for samples annealed for 45 and 60 minutes.

The removal of the chlorine content from the TA samples is further verified using grazing incidence wide-angle X-ray scattering (GIWAXS) experiments. From **Figure 5(a)-(e)**, the GIWAXS diffraction peaks located at $q \sim 1.0$ and $\sim 2.0$ Å$^{-1}$ correspond to the (110) and (220) lattice planes, in line with the XRD peaks at 14.18° and 28.6°[38,39] as shown in Figure S2. The $PbI_2$ diffraction peak at $q \sim 0.9$ Å$^{-1}$ appears with arc-like scattering with preferred out-of-plane orientation[40]. The $PbI_2$ peak intensity increases with the longer annealing time as evidenced by the XRD peaks at 12.65° also suggest. Notably, the $CH_3NH_3PbCl_3$ crystals exhibit preferred out-of-plane orientation for the 2 min and 10 min annealed TA films, as evidenced by the peak at $q \sim 1.1$ Å$^{-1}$, which further emphasizes the presence of chlorine within the TA samples. Consistent with the XRD and WDX results, the 2 min annealed films have the strongest diffraction intensity (at $q \sim 1.1$ Å$^{-1}$) corresponding to $CH_3NH_3PbCl_3$ crystals. The $CH_3NH_3PbCl_3$ phase then gradually decreases and vanishes completely for the 30 minutes and longer thermally annealed samples. To probe the evolution of phases during thermal annealing, in-situ GIWAX measurements were performed based on prepared 2-minute annealed TA perovskite films as a function of thermal annealing time as shown in **Figure 5(f)**. The $PbI_2$ gradually increases with a longer annealing process while the $CH_3NH_3PbCl_3$ phase gradually decreases and vanishes at around $\sim 100$ seconds of the 100 °C thermal annealing process.

The evolution of the *q* position of the TA perovskite (110) peak during thermal annealing can give information about the changes in the composition of TA perovskite phase (i.e. incorporation or removal of Cl), as shown in Figure S7. During the initial ramping of temperature from 25 °C to 100 °C, a change in the *q* position to lower values (increase in (110) *d*-spacing) can be attributed to lattice expansion dominated by thermal expansion. The subsequent changes in *q* position occur at a constant temperature (100 °C) and hence can be attributed to changes in the composition of the TA perovskite phase. Upon reaching 100 °C, the evolution of the *q* position indicates a shrinkage of lattice constant for the initial 5 min of annealing followed by an expansion of lattice constant for a longer annealing time. We recall that $CH_3NH_3PbCl_3$ phase dissociates during the initial annealing time giving rise to the release of Cl ions that can be incorporated into the TA perovskite phase. Therefore, we hypothesize that the initial shrinkage of the lattice constant of TA perovskite phase results from the incorporation of smaller ions (such as Cl) into the TA perovskite phase. These Cl ions are perhaps supplied from $CH_3NH_3PbCl_3$ phase dissociation. However, for longer annealing times, some of these Cl leave the film due to the high volatility of Cl, in agreement with previous studies. These results reveal that careful optimization of annealing time can be used to control the Cl content in TA perovskite films.

The indoor photovoltaic properties of the devices with different thermal annealing times are characterized to investigate the effect of chlorine release and $PbI_2$ growth. In terms of **Figure 6**, Figure S8 and Table S2, the optimized annealing time is 2 minutes, with these devices reaching the highest PCE of 26.6% and 30.1% for forward and reverse bias, respectively. The lowest PCE is obtained from 30 min-annealed devices (forward: 18.4%; reverse: 26.8%), followed by 60 min-annealed devices (forward: 19.0%; reverse: 28.9%). Regarding the variation of photovoltaic parameters, 2 min-annealed samples hold the highest $V_{OC}$ of 0.87 V and 0.92 V among the five conditions, with $V_{OC}$ is continuously decreasing during the annealing process, indicating the gradual reduction of chlorine content and narrowing of the bandgap. Notably, 2 min-annealed devices gave the highest FF of 57.2% and 78.4%. This drop in FF with longer annealing suggests the worsening of interface conditions and transport properties. The 2 min-annealed TA samples show the highest MPPT PCE as well [Figure 6(b)]. The increase in the $PbI_2$ phase with an increase in thermal annealing time can hinder the charge transport, deteriorating the PV properties. The thermal annealing study thus shows that the presence of chlorine in the TA films is contributing to better device properties.

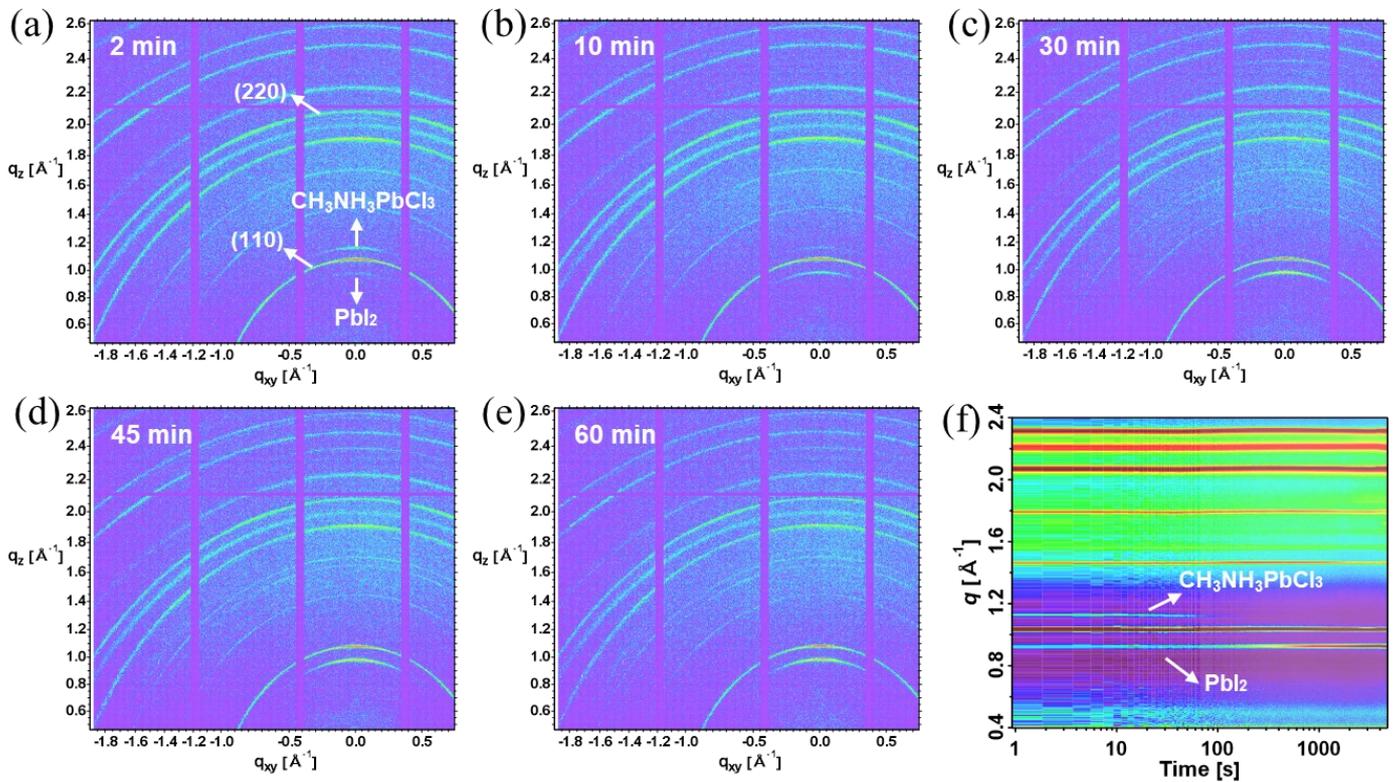

*Figure 5. (a) – (e)* Grazing incidence wide-angle X-ray scattering *diffraction patterns of TA films thermally annealed at 100 °C for different durations. With a longer thermal annealing time, the PbI$_2$ diffraction peaks increase whereas the* CH$_3$NH$_3$*PbCl$_3$ diffraction peaks decrease (f) False-color plot of in-situ GIWAXS pattern during the thermal annealing process for the pre-prepared 2 min-annealed TA film.*

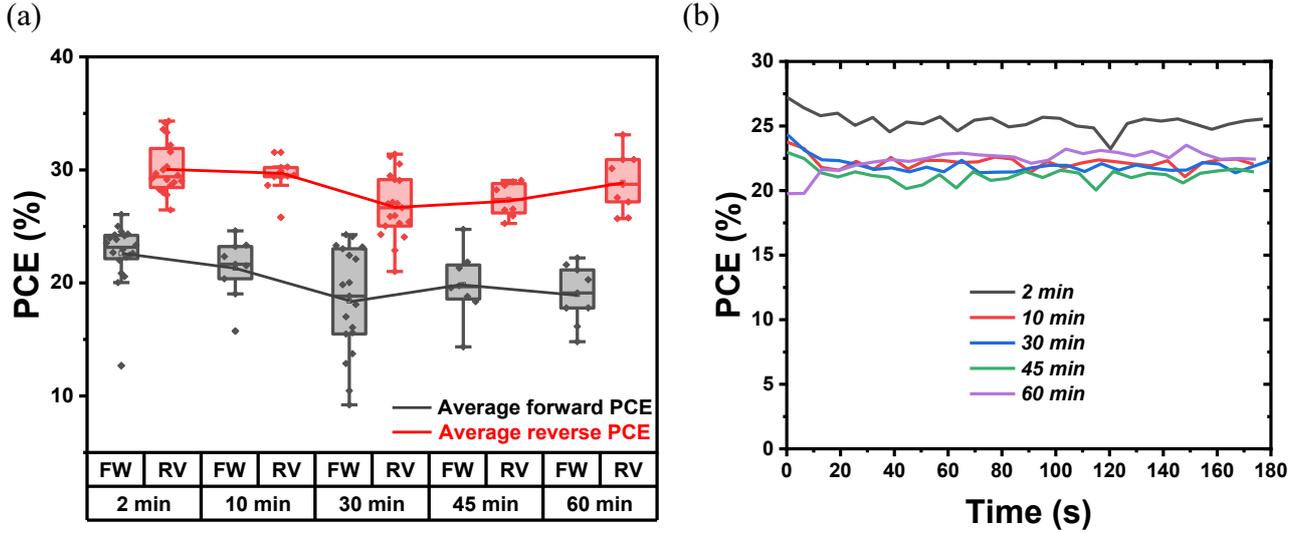

*Figure 6. (a) The statistics of PCE of devices as a function of different thermal annealing time under 1000 lux warm white LED illumination. The 2 min-annealed samples hold the highest average PCE of 26.6% and 30.1%. (b) The comparison of MPPT PCE of devices as a function of different thermal annealing time*

## 2.4. Hysteresis properties

Our recent investigation has shown that compared to 1 sun illumination, the halide perovskite photovoltaic devices demonstrate a completely different *J-V* hysteresis behaviour under indoor lighting conditions, depending on the selection of the device architecture and the photoactive layers[17]. Addressing *J-V* hysteresis is a critical aspect since reliable PCE/power output is required to self-power an external circuit using a photovoltaic device. Compared to $CH_3NH_3PbI_3$, TA devices show slightly greater *J-V* hysteresis under 1 sun illumination, [Figure 2(e)]. However, under indoor lighting illumination, *J-V* hysteresis of TA devices is suppressed in comparison to the control $CH_3NH_3PbI_3$ devices [Figure 2(b)]. The more pronounced *J-V* hysteresis of TA devices under 1 sun compared to indoor illumination can be related to the higher possibility of light-induced ion migration effects under 1 sun light intensity and the presence of different types of halide ions in the TA composition[41–45]. Previously we have shown that the $SnO_2$/perovskite interface can contribute to *J-V* hysteresis in halide perovskite indoor photovoltaic devices[17]. However, the origin of the hysteresis effect can also be due to the interfacial defects existing at the perovskite/Spiro-OMeTAD interface[46]. In this case, we employed P3HT as the hole transport layer to replace Spiro-OMeTAD. Under indoor lighting, the PCE of P3HT-based devices reached 20.4% and 20.6% for forward and reverse bias.

Notably, the hysteresis of P3HT-based devices is reduced drastically compared to Spiro-OMeTAD based devices, as shown in Figure S9, indicating that the Spiro-OMeTAD interface as well as the $SnO_2$/perovskite interface are contributing to the *J-V* hysteresis[17]. However, the photovoltaic performances of P3HT-based devices are consistently lower than Spiro-OMeTAD devices and one contributing factor could be the low conductivity of the undoped P3HT films used as HTL. More optimization is needed at the charge transporting layer/perovskite buried interfaces to eliminate the *J-V* hysteresis without compromising the PCE of the TA based indoor photovoltaic devices.

## 2.5. Powering a temperature sensor

To investigate the reliability of the TA indoor photovoltaic devices as a power source for sensors, a sensor powering experiment was carried out. Rajan *et al.* developed a wearable graphene-based textile temperature sensor with a 1 V requirement for operation[47]. The graphene sensor is comprised of three layers of graphene and coated on a single polypropylene textile fibre which is highly flexible. The graphene-based sensor shows a negative thermal coefficient of resistance, indicating that its intrinsic resistance is reduced with increased temperature. The output voltage from TA devices is tested in advance under two indoor illumination levels of 220 lux and 1000 lux. In this case, two pixels of n-i-p TA devices on a single substrate are connected in series to obtain a higher output voltage. Two illumination conditions are tested: a) ambient indoor CFL light from the ceiling (220 lux); b) domestic LED table lamp (1000 lux). The obtained output voltage from the two series connected photovoltaic devices under the two indoor illumination conditions is 1.62 V and 1.86 V, respectively. **Figure 7(a)** shows the experimental setup. The Keithley source meter is only used to measure the current across the graphene temperature sensor, no extra voltage from the Keithley source meter is applied. **Figure 7 (b)** shows the resistance variation of the sensor, powered using TA devices under the two selected indoor illumination conditions. The temperature sensor shows a consistent trend in resistance reduction with an increase in temperature, similar to that previously reported using a Keithley source meter as the voltage source[47]. These results indicate that TA perovskite-based indoor photovoltaic devices are reliable for self-powering the low-power sensors.

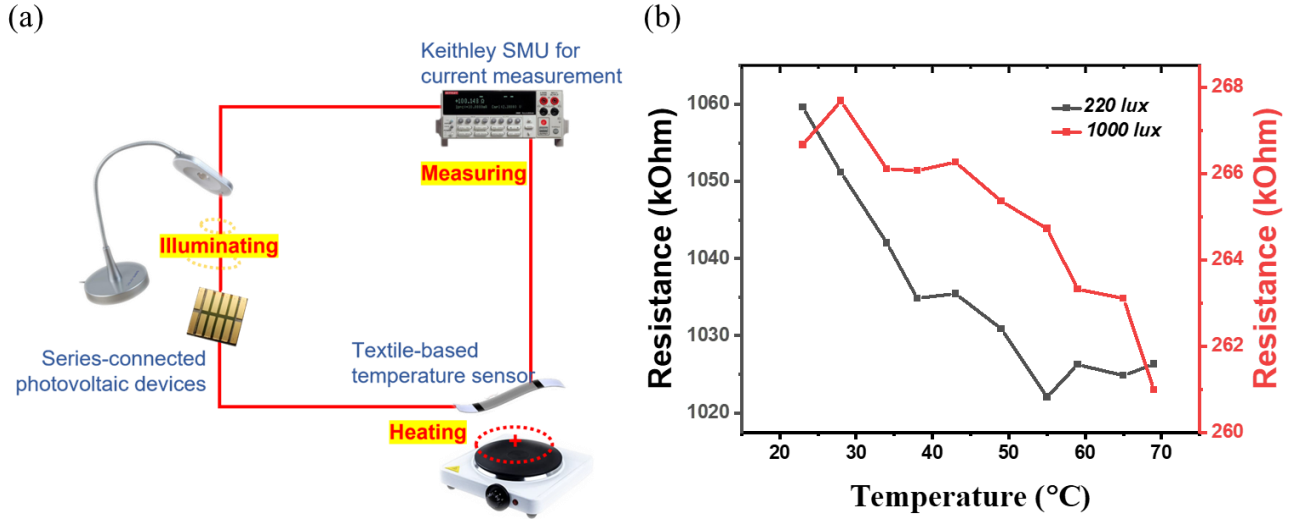

*Figure 7. (a) Illustration of sensor powering with TA perovskite indoor photovoltaic device powering a graphene-based textile temperature sensor. (b) Resistance variation of the temperature sensor in two different illumination conditions.*

## 3. Conclusion

The need for efficient indoor light harvesting was addressed by developing a composition tuned wide bandgap triple anion perovskite $CH_3NH_3PbI_{2.6}(BrCl)_{0.2}$. The corresponding indoor photovoltaic devices were capable of delivering a steady-state output power of 75.4 µW/cm$^2$ under 1000 lux warm white LED illumination and successfully self-powered a textile integrated graphene based temperature sensor. Detailed microstructural and optoelectronic investigations of the triple anion perovskite unravelled its excellent crystalline quality, widened bandgap, less density of trap states and longer carrier lifetime, all contributing positively to the enhanced photovoltaic properties. Our study revealed that to keep the Cl in the triple anion perovskite and to reap the beneficial effects of high $V_{OC}$ and enhanced charge carrier lifetime, the thermal annealing duration should be carefully optimized. Replacing the Spiro-OMeTAD hole transporting layer with P3HT was found to reduce the *J-V* hysteresis under indoor lighting conditions. The study shows the promise of simply processed triple anion perovskites for indoor photovoltaic cells to sustainably power the IoT wireless sensor components.

# Acknowledgements

LKJ acknowledges funding from UKRI-FLF through MR/T022094/1. LKJ also acknowledges, Professor Iain Baikie for the work function and APS measurements and Professor Phil King and Gordon Kentish, School of Physics and Astronomy, University of St Andrews for the XRD measurements and would like to acknowledge (EPSRC): EP/T023449/1. This research used resources of the Advanced Light Source, a U.S. DOE Office of Science User Facility under contract no. DE-AC02-05CH11231. Work was performed at beamline 12.3.2, beamline scientist Nobumichi Tamura.

# Conflict of Interest

The authors declare no conflict of interest.

# Data accessibility

The research data underpinning this publication can be accessed at
https://doi.org/10.17630/93ab4fa0-34e3-45a9-aa48-032dec3f675e [REF]